\documentclass[twocolumn,superscriptaddress,letter]{revtex4}%
\usepackage{amssymb}
\usepackage{color}
\usepackage{graphicx}
\usepackage{dcolumn}
\usepackage{bm}
\usepackage[header,title,page,titletoc]{appendix}
\usepackage{amsmath}
\usepackage{subfigure}
\usepackage{amsfonts}
\usepackage[colorlinks, citecolor=red]{hyperref}%
\providecommand{\U}[1]{\protect \rule{.1in}{.1in}}
\begin{document}
\title{Tunable Weyl Semi-metal and its Possible Realization in Optical Lattice}
\author{Xiao Kong}
\author{Ying Liang}
\author{Su-Peng Kou}
\email{spkou@bnu.edu.cn}
\affiliation{Department of Physics, Beijing Normal University.}

\begin{abstract}
Weyl semimetal (WSM) is an exotic topological state in condensed matter
physics. In this paper, based on a two-band cubic lattice model, we studied
WSMs with a pair of tunable Weyl nodes. It is pointed out that there exist
three types of WSMs with different tilt strengths: type-I WSM, type-II WSM
and type-1.5 WSM. In particular, type-1.5 WSM
has one type-I node and one type-II node. We studied chiral modes, surface Fermi arcs and quantum
anomalous Hall effect in different types of WSMs. In addition, we give an experimental
setup to realize the different types of WSMs based on timely technique.\ 

\begin{description}
\item[PACS numbers] 
37.10.Jk, 72.90.+y, 67.85.Lm
\end{description}
\end{abstract}
\maketitle

Weyl semimetal (WSM) is a three-dimensional Graphene-like system with
low-energy relativistic excitations\cite{prb83205101,
naturecom67373,prx5031013,naturephy11724}. In condensed matter physics, WSM
was proposed to separate a single Dirac node into two Weyl nodes by breaking
either time reversal symmetry or inversion symmetry. Thus, Weyl nodes always
appear and disappear in pairs with opposite chiralities. For example, there
are 12 pairs of Weyl nodes in the pyrochlore iridates $\mathrm{A}%
_{2}\mathrm{Ir}_{2}\mathrm{O}_{7}$ and in the $\mathrm{TaAs}$. Recently, a new
type of Weyl semimetal is proposed, which is called type-II Weyl semimetal (we
denote it by WSM-II and traditional WSM by WSM-I)\cite{nature527495}. For
WSM-II, the electron and hole pockets touch and the dispersions become
anisotropic\cite{arxiv160100890,arxiv160404030, arxiv160408457}. WSM-II was
predicted in systems of $\mathrm{Mo}_{x}\mathrm{W}_{1-x}\mathrm{Te}_{2}%
$\cite{naturecom710639}, and then the surface Fermi arcs in $\mathrm{MoTe}%
_{2}$ were observed\cite{arxiv160402116}. Due to its novel physics properties,
such as chiral magnetic effect, negative magnetoresistance effect and surface
Fermi arcs, WSM becomes hot topic in condensed matter
physics\cite{prb84075129,prx5031023, arxiv13111506,prb86115133,
prb88104412,prb88115119,prb88245107, prl111027201,prb87235306}. Now, searching
for materials of WSMs with few Weyl nodes is still an open problem.

The studies of ultracold atoms in optical lattices are extensively
developed\cite{Greiner,jak}. Because of their precise control over the system
parameters and defect-free properties, ultracold atoms in optical lattices
provide an ideal platform to study many-body physics in condensed
matters\cite{Lewenstein,blo}. Since the simulating magnetic fields (especially
the non-Abelian ones) in ultra-cold atom gases was achieved, investigations on
topological phases become an important issue. In Ref.\cite{prl114225301},
WSM-I was designed on optical lattices. However, WSM-II has not been simulated
on optical lattices of ultracold atoms.

In this paper, based on an experimental setup in optical lattice for generating and freely
controlling WSM, we realize different types of WSMs with a pair of tunable
Weyl nodes. In particular, we realize a new region of WSM which contains one
type-I node and one type-II node. We call it type-1.5 WSM (WSM-1.5). In this
paper, we will show the physical properties for different types of WSMs and
discuss how to realize them.

\textit{Model Hamiltonian of generalized WSMs: }Our starting point is a
two-band cubic lattice model for WSM-I\cite{prb84075129}, of which the
Hamiltonian is written as
\begin{align}
H_{I}  &  =\sum_{i}t_{x}(c_{i\uparrow}^{\dagger}c_{i+\widehat
{x}\downarrow}+c_{i\downarrow}^{\dagger}c_{i+\widehat{x}\uparrow}%
)-t_{y}(c_{i\uparrow}^{\dagger}c_{i+\widehat{y}\downarrow}\nonumber \\
&  -c_{i\downarrow}^{\dagger}c_{i+\widehat{y}\uparrow})-\frac{m}%
{2}(c_{i\uparrow}^{\dagger}c_{i+\widehat{y}\downarrow}+c_{i\downarrow
}^{\dagger}c_{i+\widehat{y}\uparrow})\nonumber \\
&  -e^{i\pi/2}t_{z}(c_{i\uparrow}^{\dagger}c_{i+\widehat{z}\uparrow
}-c_{i\downarrow}^{\dagger}c_{i+\widehat{z}\downarrow})\nonumber \\
&  -\frac{m}{2}(c_{i\uparrow}^{\dagger}c_{i+\widehat{z}\downarrow
}+c_{i\downarrow}^{\dagger}c_{i+\widehat{z}\uparrow})+h.c.\nonumber \\
&  +2(m-\cos k_{0})(c_{i\uparrow}^{\dagger}c_{i\downarrow}+c_{i\downarrow
}^{\dagger}c_{i\uparrow}), \label{h0real}%
\end{align}
where $i$ denotes the lattice site, $c$ and $c^{\dagger}$ are annihilation and
creation operators, $\widehat{x}$, $\widehat{y}$ and $\widehat{z}$ represent
the nearest-neighbor vectors along $x$, $y$ and $z$ axes and $h.c.$ is short
of hermitian conjugate. $t_{x,y,z}$ are hopping parameters and $m$ is the
strength of on-site "external field". $k_{0}$ denotes a constant wave vector.
$\uparrow$ and $\downarrow$ are pseudo spin degrees of freedom. By Fourier
transformation, Eq.\ref{h0real} turns into $H_{I}=\sum_{k}c_{k}%
^{\dagger}\mathbf{h}_{I}(k)c_{k},$ where%
\begin{align}
\mathbf{h}_{I}(k)  &  =[2t_{x}(\cos k_{x}-\cos k_{0})\nonumber \\
&  +m(2-\cos k_{y}-\cos k_{z})]\cdot \sigma_{x}\nonumber \\
&  +2t_{y}\sin k_{y}\cdot \sigma_{y}+2t_{z}\sin k_{z}\cdot \sigma_{z},
\label{h01}%
\end{align}
where $\sigma_{x,y,z}$ represent Pauli matrices. For the case of
$m>2t_{x}(1+\cos k_{0})$, this model has two Weyl nodes at $\mathbf{K}%
_{+}=(k_{0},$ $0,$ $0)$ and $\mathbf{K}_{-}=(-k_{0},$ $0,$ $0)$, with Fermi
velocities $\mathbf{v}_{+}=(-2t_{x}\sin k_{0},$ $2t_{y},$ $2t_{z})$ and
$\mathbf{v}_{-}=(2t_{x}\sin k_{0},$ $2t_{y},$ $2t_{z})$. The low-energy
Hamiltonians near nodes can be written as $\mathbf{h}_{I}(k)\simeq%
%TCIMACRO{\dsum \limits_{n=+,-,i}}%
%BeginExpansion
{\displaystyle \sum \limits_{n=+,-,i}}
%EndExpansion
v_{n,i}k_{i}\cdot \sigma_{i}$. The chiralities of nodes are $\mathrm{sgn}%
(\prod_{i}v_{n,i})$ which could be $\pm1$.

Based on the WSM-I Hamiltonian in Eq.\ref{h01}, we obtain a generalized WSM by
tilting the two nodes. In general, to arbitrarily tune the Weyl nodes, we may
add the following terms $\mathbf{h}_{\mathrm{tilt}}=\mathbf{h}_{ox}%
+\mathbf{h}_{sx}+\mathbf{h}_{sy}+\mathbf{h}_{sz}+\mathbf{h}_{oy}%
+\mathbf{h}_{oz}$ where%
\begin{align}
\mathbf{h}_{ox}  &  =\sum_{i}v_{ox}(c_{i\uparrow}^{\dagger}c_{i+\widehat
{x}\uparrow}+c_{i\downarrow}^{\dagger}c_{i+\widehat{x}\downarrow
})+h.c.\nonumber \\
&  -2v_{ox}\cos k_{0}(c_{i\uparrow}^{\dagger}c_{i\uparrow}+c_{i\downarrow
}^{\dagger}c_{i\downarrow}),\nonumber \\
\mathbf{h}_{sx}  &  =\sum_{i}e^{i\pi/2}v_{sx}(c_{i\uparrow}^{\dagger
}c_{i+\widehat{x}\uparrow}+c_{i\downarrow}^{\dagger}c_{i+\widehat{x}%
\downarrow})+h.c.\nonumber \\
&  -2v_{sx}\sin k_{0}(c_{i\uparrow}^{\dagger}c_{i\uparrow}+c_{i\downarrow
}^{\dagger}c_{i\downarrow}),\nonumber \\
\mathbf{h}_{sy/sz}  &  =\sum_{i}e^{i\pi/2}v_{sy/sz}(c_{i\uparrow}^{\dagger
}c_{i+\widehat{y}/\widehat{z}\uparrow}+c_{i\downarrow}^{\dagger}%
c_{i+\widehat{y}/\widehat{z}\downarrow})+h.c.,\nonumber \\
\mathbf{h}_{oy/oz}  &  =\sum_{i}v_{oy/oz}(c_{i\uparrow}^{\dagger}%
c_{i+\widehat{y}/\widehat{z}+\widehat{x}\uparrow}+c_{i\downarrow}^{\dagger
}c_{i+\widehat{y}/\widehat{z}+\widehat{x}\downarrow})\nonumber \\
&  -\sum_{i}v_{oy/oz}(c_{i\uparrow}^{\dagger}c_{i+\widehat{y}/\widehat
{z}-\widehat{x}\uparrow}+c_{i\downarrow}^{\dagger}c_{i+\widehat{y}/\widehat
{z}-\widehat{x}\downarrow})+h.c., \label{hoz1}%
\end{align}
where $\mathbf{I}$ is $2\times2$ identity matrix. Thus, the total low-energy
effective model becomes
\begin{align}
\mathbf{h}_{\mathrm{total}}(k)  &  =\mathbf{h}_{I}(k)+\mathbf{h}%
_{\mathrm{tilt}}\simeq v_{F}[\alpha_{\pm}\mathbf{n_{\pm}}\cdot(\mathbf{k}%
-\mathbf{K_{\pm}})I\nonumber \\
&  \pm(\mathbf{k}-\mathbf{K_{\pm}})\cdot \mathbf{\sigma}\pm \dfrac{b_{\pm}}%
{2}I], \label{h2}%
\end{align}
where $\mathbf{n_{\pm}}$ is tilt strength vector which defines tilt direction
of nodes, $\alpha_{\pm}$ is tilt strength, a parameter describing amplitude of
tilt and $b_{\pm}$ the chemical potential of nodes.

\begin{figure}[ptb]
\begin{center}
\includegraphics[clip,width=0.5\textwidth]{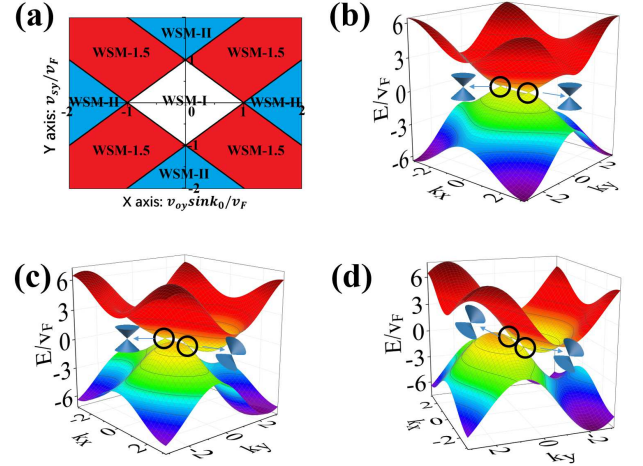}
\end{center}
\caption{(color online) Illustration of different WSMs: (a) phase diagram of
parameters $v_{oy}$ and $v_{sy}$; (b) WSM-I without tilts of nodes; (c)
WSM-1.5 with one type-I node and one type-II node, parameters are setted as
$\mathbf{n}_{+}=(0,1,0) $, $\alpha_{+}=-1.2$ and $\alpha_{-}=0$; (d) WSM-II
with same tilts perpendicular to the node-separation, parameters are setted as
$\mathbf{n}_{+}=\mathbf{n}_{-}=(0,1,0) $ and $\alpha_{+}=\alpha_{-}=-1.2$.}%
\label{scheme}%
\end{figure}

We have four free parameters that describe the low energy physics of a node,
$\mathbf{K}_{+}$ $($or $\mathbf{K}_{-})$, $\mathbf{n_{+}}$ (or $\mathbf{n_{-}%
}$), $\alpha_{+}$\ (or $\alpha_{-}$), $b_{+}$ ($b_{-}$). A detailed
calculations are given in supplementary-materials. For a given Weyl node,
there exists two phases: in the region of $\alpha_{\pm}<1$, the (tilting) Weyl
node belongs to type-I; in the region of $\alpha_{\pm}>1,$ the (tilting) Weyl
node belongs to type-II, in which the electron pocket and hole pocket contact
each other. As a result, there exist three types of WSMs: for the case of
$\alpha_{+}<1$ and $\alpha_{-}<1$ we call the WSM to be type-I WSM (WSM-I);
for the case of $\alpha_{+}>1$ and $\alpha_{-}>1$, we call it type-II WSM
(WSM-II); for the case of $\alpha_{+}>1$ and $\alpha_{-}<1$ and the case of
$\alpha_{+}<1$ and $\alpha_{-}>1$, we call it type-1.5 WSM (WSM-1.5). We show
the phase diagram and dispersions of WSMs in Fig.\ref{scheme}.

\begin{figure}[h]
\begin{center}
\includegraphics*
[width=0.5\textwidth]{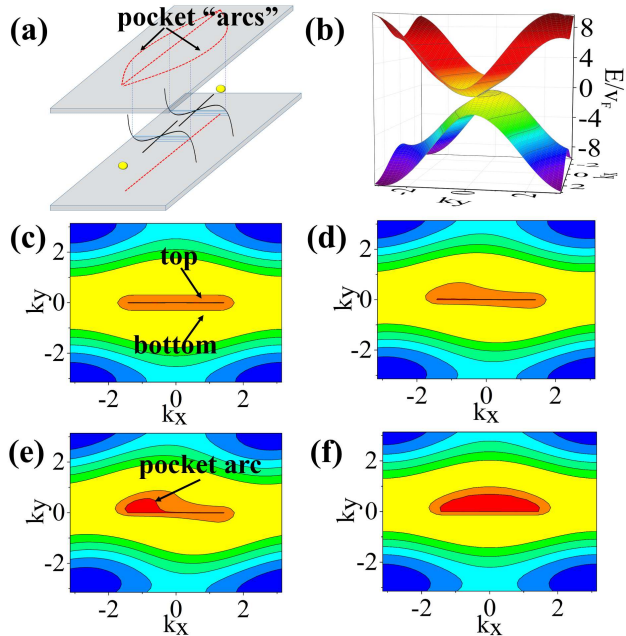}
\end{center}
\caption{(color online) Surface Fermi arcs. (a) A scheme of Fermi arcs in
WSM-II, surface states are tilted so that their electron and hole pockets also
cross Fermi surface. There is an additional Fermi circle appears in one
surface. We split them into two arcs and call them pocket arcs. The yellow
balls are Weyl points and the red dash lines are Fermi arcs; (b) Type-II
surface states; (c)-(f) show energy contours of surface state. The edge of
orange area is cut by $E=-0.3v_{F}$ and the edge of red area is cut by
$E=-0.015,$ respectively. The parameters are $\alpha_{+}=\alpha_{-}=0$ for
(c),  $\alpha_{+}=0,\alpha_{-}=0.5$ for (d), $\alpha_{+}=0,\alpha_{-}=1.2$ for
(e) and $\alpha_{+}=\alpha_{-}=1.2$ for (f). The tilting direction is
$\mathbf{n}_{+}=\mathbf{n}_{-}=(0,1,0)$. }%
\label{yarcs}%
\end{figure}

\textit{Surface Fermi arcs for different types of WSMs:} In our model, nodes
are separated along $\hat{x}$-direction. Fermi arcs appear except for the
projection along $\hat{x}$-direction. For WSM-I with $\alpha_{\pm}=0$, Fermi
arcs on opposite surfaces have the same dispersion. See the results in
Fig.\ref{yarcs}(c). To study the surface states in a WSM system, we cut the
3-dimensional system to parallel 2-dimensional (2D) systems piece by piece
along $\hat{x}$-direction and get a 2D topological insulator with $k_{0}%
>k_{x}>-k_{0}$, of which the surface states have the effective Hamiltonian as
$\mathbf{h}_{\mathrm{surface}}=2t_{y}\sin k_{y}\sigma_{z},$ $k_{x}\in(-k_{0},$
$k_{0})$ where $\sigma_{z}$ is the pseudo-spin operator of surfaces.

We then study the surface Fermi arcs in WSMs with $\alpha_{\pm}\neq0$.
Firstly, we study the tilting effect on surface states along $\hat{x}%
$-direction (the tilt direction of node $(k_{0},0,0)$ is set to be
$\mathbf{n}_{+}=(1,0,0)$). Due to the existence of an additional term
$\mathbf{h}_{\mathrm{tilt}}=v_{F}\alpha_{+}l^{\prime}\cdot \mathbf{I}$ on each
$k_{y}$-$k_{z}$ plane where $l^{\prime}=k_{x}-k_{0} $, the effective
Hamiltonian of the surface states $\mathbf{h}_{\mathrm{surface}}$ has an
energy shift that is proportional to $l^{\prime}$. As a result, the shape of
Fermi arcs changes. However, the situation changes when we study the tilting
effect on surface states along $\hat{y}$-direction. Now, the tilt direction of
node $(k_{0},0,0)$ becomes $\mathbf{n}_{+}=(0,1,0)$. Because the translation
invariance along $\hat{y}$-direction is not broken, we can easily derive the
surface effective Hamiltonian by the method in Ref.\cite{shenshunqing} and
obtain $\mathbf{h}_{\mathrm{surface}}=v_{F}k_{y}\sigma_{z}+v_{F}\alpha
_{+}k_{y}\mathbf{I}$. As a result, the dispersion of surface states near nodes
are tilted and shows similar behaviour as those in bulk. For example, in a
WSM-II, the surface states are also in "type-II". See the illustration in
Fig.\ref{yarcs}.

\textit{Chiral modes of different types of WSMs in magnetic field: }When
applying magnetic field $B\hat{x}$ to WSM-I system, there are two chiral
modes: one comes from the zeroth Landau level near node $k_{0}$, the other
comes from the zeroth Landau level near node $-k_{0}$. As a result,WSM-I
behaves chiral anomaly such as negative magnetoresistance effect and chiral
magnetic effect\cite{prb86115133,prb87235306,crhy14857}.

\begin{figure}[h]
\begin{center}
\includegraphics*
[width=0.5\textwidth]{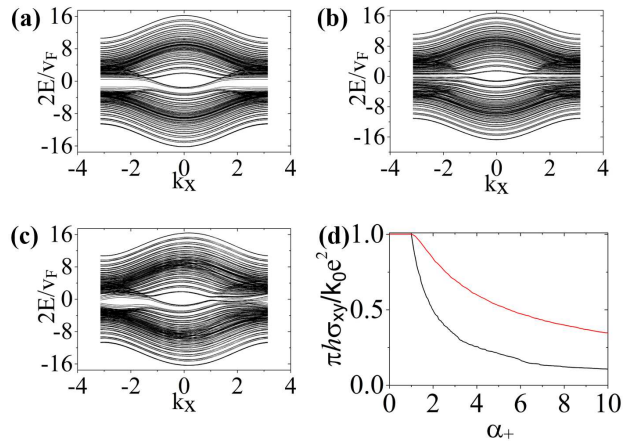}
\end{center}
\caption{(color online) (a) Dispersion of WSM-I in a magnetic field along
$\hat{x}$. There are two chiral modes near Fermi surface $E=0$; (b) Dispersion
of WSM-II with $\mathbf{n}_{+}=\mathbf{n}_{-}=(0,1,0)$ and $\alpha_{+}%
=\alpha_{-}=1.2$ in a magnetic field along $\hat{x}$. There are no chiral mode
and the spectrum is gapped; (c) Dispersion of WSM-1.5 with $\alpha_{-}=0$,
$\mathbf{n}_{+}=(0,1,0)$ and $\alpha_{+}=1.2$ in a magnetic field along
$\hat{x}.$ There exists a chiral mode; (d) Hall conductivity in WSMs. The
black line is the result for the case of $\mathbf{n}_{+}=\mathbf{n}%
_{-}=(0,1,0)$ and $\alpha_{+}=\alpha_{-}=\alpha$. The red line is the result
for the case of $\mathbf{n}_{+}=(0,1,0)$, $\alpha_{+}=\alpha$ and $\alpha
_{-}=0$.}%
\label{Landaulevels}%
\end{figure}

We then study the chiral modes in WSMs with $\alpha_{\pm}\neq0$. According to
Ref.\cite{arxiv160404030}, when a magnetic field perpendicular to $\mathbf{n}$
is applying to the system, the Landau levels in WSM-IIs are collapsing, the
system then have no chiral modes\cite{nature527495}. Thus, when magnetic field
is perpendicular to tilting direction, no chiral mode appears and the system
doesn't behave chiral anomaly. For the case of WSM-1.5, one node is type-I and
the other is type-II. In Fig.\ref{Landaulevels}, we calculate the energy
spectrum of WSM-1.5 in magnetic field. Obviously, there is a linear chiral
mode near the type-I node $(-k_{0},0,0)$ and the bands near the type-II node
$(k_{0},0,0)$ are gapped.

In addition, we calculate the Hall conductance in different WSMs and show the
results in Fig.\ref{Landaulevels}(d). For WSM-I with zero chemical potential,
there exists quantum anomalous Hall effect and the Hall conductance is
$\sigma_{yz}=\dfrac{e^{2}k_{0}}{2\pi^{2}\hslash}$ (that corresponds to the
platform in Fig.\ref{Landaulevels}(d)). We then tilt the WSM with zero
chemical potential into WSM-II and WSM-1.5. Due to finite density at Fermi
surface, the Hall conductance will not be a constant and decrease with tilt strength.

\textit{Experimental setup for different types of WSMs in optical lattice: }Designing an experimental setup in optical lattice, which possesses both tunable
anisotropic properties and intact band structure, is impending for the studies
on WSM-II and WSM-1.5. In this part, we use Fermi atom gases with two
long-life hyperfine levels to simulate different types of
WSMs\cite{njp14015007}.

At the beginning, we construct the optical lattice by standing waves:
$V(x,y,z)=-V_{0}\sum_{i=x,y,z}[\cos^{2}(qi)-V_{0}^{\prime}\cos^{2}(2qi)],$
where $x,$ $y,$ $z$ denote the different directions, $q=\frac{2\pi}{\lambda}$
is the recoiling momentum and $V_{0}$ and $V_{0}^{\prime}$ are positive
potential amplitudes. The standing waves have a cubic array of minima
separated by second minima sites. The atoms are trapped in the standing waves and
we can use the Wannier functions $\omega_{n,\mathbf{R}}(\mathbf{r})$ to
describe the states on each site, where $n$ is band index, $\mathbf{R}$ the
location of site and $\mathbf{r}$ the displacement from $\mathbf{R}$. Let us
consider $^{40}\mathrm{K}$ atoms in the present of a uniform magnetic field
$\mathcal{B}$, and employ the two hyperfine manifolds of ground states:
$F=9/2$ and $F=7/2$. The energies are $E_{9/2,m_{F}}=g_{F}\mu_{B}%
\mathcal{B}m_{F}$ and $E_{7/2,m_{F}}=\Delta_{HF}-g_{F}\mu_{B}\mathcal{B}m_{F}%
$, where $g_{F}$ is the hyperfine Land\'{e} factor, $\mu_{B}$ the Bohr
magneton and $m_{F}$ the projection along magnetic field.

Now, setting large potential amplitudes, we get a cubic optical lattice with
atoms trapped in minima and second minima sites without any hopping. To
realize the desired hopping terms, we apply four Raman laser beams to the
system. We take the following term as an example: $-t_{x}e^{i\phi_{a}}%
c_{a}^{\dagger}\sigma_{x}c_{a+\widehat{x}}+h.c.$, where $t_{x}$ is the real
hopping strength, $a$ the site of optical lattice (that is a local minimum),
$\widehat{x}$ the vector from a site to its nearest site along $x$ and
$\phi_{a}$ the phase of hopping; $c_{a}$ is annihilation operators with
(pseudo) spins $c_{a}=(c_{a\uparrow},c_{a\downarrow})^{T}$ and $\sigma_{x}$
the Pauli matrix. This term corresponds to the first hopping term in
Eq.\ref{h0real} for the case of $\phi_{a}=2\pi \ast \mathrm{integer}$.

\begin{figure}[h]
\begin{center}
\includegraphics*
[width=0.5\textwidth]{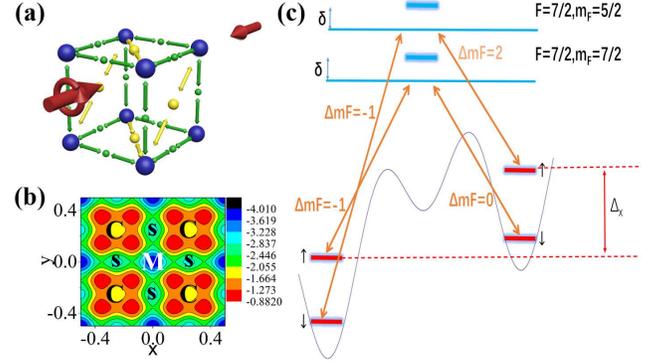}
\end{center}
\caption{(color online) (a) An illustration of our purpose to realize WSMs.
Blue balls represent minima sites and green and yellow balls represent second
minima sites. Double sided arrows denote the Raman couplings. Red arrows
denote Raman lasers for a Raman coupling with $\Delta m_{F}=-1$, of which a
circle arrow marks an unit of angular momentum carried by a $\sigma$ laser;
(b) The lattice potential for all hyperfine states without gradient potential.
\textbf{M }denote minima sites, \textbf{S }denotes second minima sites and
\textbf{C} center second minima sites; (c) A scheme of realizing spin-flipping
terms in the present of a gradient potential. }%
\label{ramanscheme}%
\end{figure}

We then employ a pair of Raman laser beams which couple a ground state to an
auxiliary state with net momentum $\mathbf{q}$ and frequency $\omega_{1},$ and
another pair of $\mathbf{q},$ $\omega_{2}$ couple auxiliary state to another
ground state. For the Wannier state $\left \vert F=9/2,m_{F}=9/2\right \rangle
_{a}$ on site $a$ and $\left \vert F=9/2,m_{F}=7/2\right \rangle _{a+\widehat
{x}}$ on site $a+\widehat{x}$, we have the energies of them $E_{a,9/2}%
=9g_{F}\mu_{B}\mathcal{B}/2,$ $E_{a+\widehat{x},7/2}=7g_{F}\mu_{B}%
\mathcal{B}/2.$ The auxiliary state $\left \vert F=7/2,m_{F}=7/2\right \rangle
_{a+\widehat{x}/2}$ on second minimum site has energy $E_{a+\widehat{x}%
/2,7/2}=\Delta_{HF}+V_{0}-7g_{F}\mu_{B}\mathcal{B}/2.$ We set the frequency
difference $\omega_{2}-\omega_{1}\approx g_{F}\mu_{B}\mathcal{B}$ so that the
ground dressed states are nearly degenerate. Moreover, for the detuning
$\delta$ to be $\delta=\Delta_{HF}+V_{0}-7g_{F}\mu_{B}\mathcal{B}-\omega_{2},$
the excited dressed states have higher energies and can be treated as
auxiliary states. By adiabatic eliminating the auxiliary states, we could
derive an effective Hamiltonian as $H_{eff}=-\frac{|\Omega|^{2}}{\delta
}e^{2i\mathbf{q}\cdot \mathbf{a}}c_{a,\uparrow}^{\dagger}c_{a+\widehat
{x},\downarrow},$ where $\Omega$ is the strength of Raman coupling between the
ground states and the auxiliary state, $\mathbf{a}$ denotes the position of
site $a$. The spin indices $\uparrow$ for hyperfine levels represents the
state $\left \vert F=9/2,m_{F}=9/2\right \rangle $ and $\downarrow$ for
$\left \vert F=9/2,m_{F}=7/2\right \rangle $. By considering another two pairs
of Raman lasers that couple the ground states to the auxiliary state
$\left \vert F=7/2,m_{F}=5/2\right \rangle _{a+\widehat{x}/2}$, we may derive
the following coupling $H_{eff}=-\frac{|\Omega|^{2}}{\delta
}e^{2i\mathbf{q}^{\prime}\cdot \mathbf{a}}c_{a,\downarrow}^{\dagger
}c_{a+\widehat{x},\uparrow}.$ As a result, the effective Hamiltonian
$-t_{x}e^{i\phi_{a}}c_{a}^{\dagger}\sigma_{x}c_{a+\widehat{x}}+h.c.$ is
obtained after fixing the parameters $t_{x}=\frac{|\Omega|^{2}}{\delta}$ and
$\phi_{a}=e^{i2\mathbf{q}\cdot \mathbf{a}}$. Therefore, the lattice model of
WSM-II in Eq.\ref{h2} could be realized by similar method step-by-step. For
example, the next-nearest-neighbor hopping could be realized by employing
auxiliary states on second minima sites in the center of four minima sites,
which is shown in Fig.\ref{ramanscheme}.

However, if we consider the next-nearest neighbor hopping (for example,
$\mathbf{h}_{oy}$) by employing two pairs of Raman lasers with the momentum
difference $\Delta \mathbf{q=(}\frac{\pi}{2|\widehat{x}|},\frac{\pi}%
{2|\widehat{x}|},0\mathbf{)}$, the lasers could also induce an additional
hopping term along $x$ and $y$ directions with $\pi$ flux per plaquette. To
avoid the interferences between different Raman processes, we apply an
anisotropy gradient potential on the system. If the potential on site $a$ is
$0$, it on $a+\widehat{x}(\widehat{y},\widehat{z})$ is $\Delta_{x}(\Delta
_{y},\Delta_{z})$ and $\Delta_{i}\neq \Delta_{j}$ if $i\neq j$. Thus, two pairs
of Raman lasers would have frequency difference $\omega_{2}-\omega_{1}%
=(\Delta_{x}+\Delta_{y}(\Delta_{z}))$ for hopping between same hyperfine state
$\left \vert F=9/2,m_{F}\right \rangle _{a}$ and $\left \vert F=9/2,m_{F}%
\right \rangle _{a+\widehat{x}+\widehat{y}(\widehat{z})}$ on $x-y(z)$ plane.
Frequency differences of other Raman lasers are also modified to match the
gradient. In the end, after adding the gradient magnetic field, a tunable
WSM-II system is designed on optical lattice. The second minima sites (or the
light potential $V_{0}^{\prime}$) is not necessary for simulating a WSM, so
the setup could be simplified as we only need one pair of Raman lasers to
induce a hopping term. However, we prefer the second minima sites to exist
because it gives an additional parameter $\delta$ to simulate more complicate phenomena.

In the end of this paper, we give a summary. We proposed a tunable two-band
lattice model in optical lattice which corresponds to a WSM system. There exist
three types of WSMs with different tilt strengthes: WSM-I, WSM-II and WSM-1.5.
We studied chiral modes, surface Fermi arcs and quantum anomalous Hall effect
in different types of WSMs. In a WSM-1.5 only one chiral mode occurs, which
may imply novel chiral anomaly in this system. The Dirac cone of surface
states are also tilted when we tilt nodes along certain directions ($\hat{y}$
or $\hat{z}$ in our model) in bulk. In addition, we designed an experimental
setup for different types of WSMs in optical lattice that provides a platform
for further studies on WSM systems.

\begin{center}
{\textbf{* * *}}
\end{center}

This work is supported by National Basic Research Program of China (973
Program) under the grant No. 2012CB921704 and NSFC Grant No. 11174035,
11474025, 11404090, We also acknowledge the support from the Fundamental
Research Funds for the Center Universities with No. 2014KJJCB26 (Y. Liang).

After we finished this work, we found the concept of type-1.5 WSM also to be proposed as hybrid WSM in Ref.\cite{arxiv160708474} a few days ago. Our works are independent to each other.

\end{document}